\pdfoutput=1
\pdfoutput=1
\pdfoutput=1
\pdfoutput=1
\pdfoutput=1
\documentclass
[a4paper]{article}

\pdfoutput=1
\pdfoutput=1
\pdfoutput=1
\pdfoutput=1
\pdfoutput=1

\usepackage[T1]{fontenc}
\usepackage{amsfonts,amssymb,amsmath} 
\usepackage{graphics,graphicx,epsfig}
\usepackage{bm}
\usepackage{color}

\usepackage[cp1251]{inputenc}
\usepackage [ english] { babel }

\usepackage{amscd}
\usepackage{amsmath}

\usepackage{xr-hyper}
\usepackage[unicode,pagebackref]{hyperref}
\hypersetup{pdftex}
\usepackage{makeidx}

\newcommand{\beq}{\begin{equation}}
\newcommand{\eeq}{\end{equation}}
\newcommand{\beqa}{\begin{eqnarray}}
\newcommand{\eeqa}{\end{eqnarray}}

\usepackage{fancyhdr}
\fancypagestyle{plain}{%
    \fancyhead{}                      

}
\newcounter{zadacha}

  \title{A binary noisy channel to model errors in printing process}

       \author{V.N. Gorbachev, E.S. Yakovleva \footnote{email: 2305lena@mail.ru}}

\date{
 \footnotesize
\emph{North-Western Institute of Printing,
\\
St.-Petersburg State University of Technology and Design,
\\
13, Djambula, St-Petersburg, 191180,
Russia}  }

\begin{document}


\maketitle
\tableofcontents
\begin{abstract}
%

To model printing noise a binary noisy channel and a set of controlled gates are introduced. The channel input is an image created by a halftoning algorithm and its output is the printed picture.
Using this channel robustness  to noise between halftoning algorithms can be studied. We introduced relative entropy to describe immunity of the algorithm to noise and tested several halftoning algorithms.   \end{abstract}

\section{Introduction}
Printing is a complicated process that can be considered from the numerous points of view.
In literature it is payed more attention to halftoning, a process that creates a binary image to reproduce it as a continuous-tone picture by a binary printing device \cite{1}.
The goal of  all halftoning techniques is to generate an image  that is perceptually similar to the original.
To estimate the original and printed image, a large number of distortion measures and criteria based on the models of the human vision system have been proposed \cite{2}.  Absence of the universal solutions stimulates progress in these fields.

In this paper we consider a printing process as a communication channel that transforms a  halftoned image into a printed image. We are interested in the main question  can the noisy channel be suitable to describe errors in printing process?  We introduced a binary channel, well known in theory of information \cite{Shannon}, and a set of controlled-gates to model  printing noise. We found that
such simple noise of the binary channel as bit-flip and erasing noise introduce a visible distortion. The introduced noisy channel allows us to consider  robustness of halftone images against to noise. By this way robustness between halftoning algorithms can be studied. Immunity to noise is important because it provides the quality of the printed images. Noise has a physical nature and results in errors in printing always. However one can reduce errors if one  chooses a halftoning algorithm such that its immunity to noise is large.

\section{Binary channel for printing}
 Any printing device has binary output, if a dot is printed or not.  From the point of view of the  theory of information the printing process can be considered as  transmitting information trough a binary channel with noise that can model errors of printing.

\begin{figure}
  \includegraphics[width=18cm]{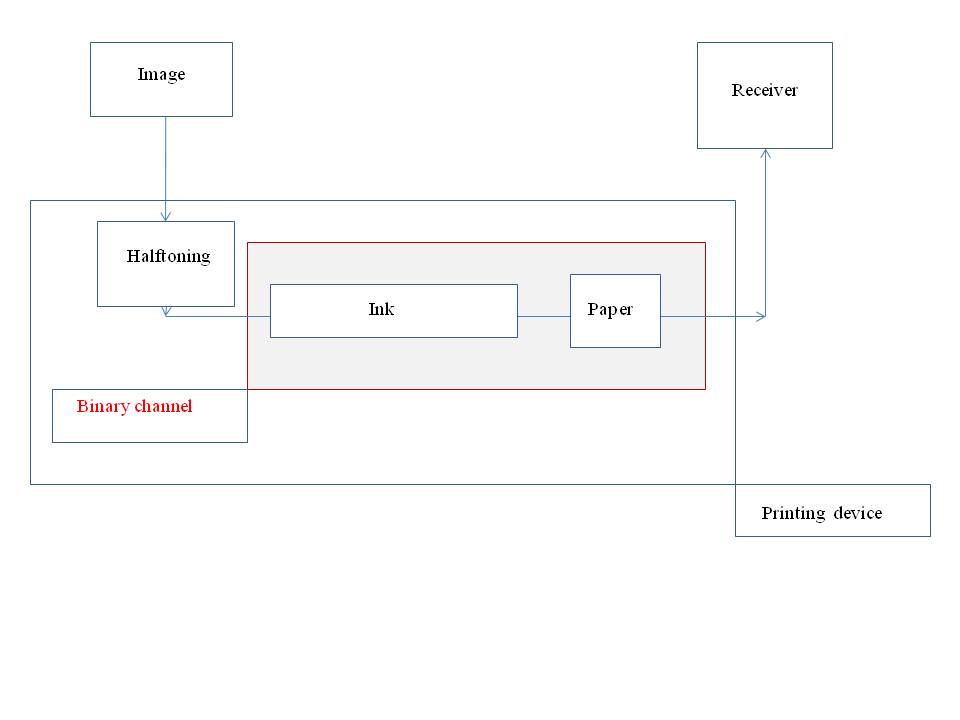}\\
  \caption{Binary channel for printing. Input halftoning image is transmitted trough the channel, a printing device, which output is a printed image.
   }\label{b1}
\hyperlink{b1}{W}
\end{figure}

To introduce the binary channel for printing, we begin with the general scheme of Shannon \cite{Shannon} for  transmitting  and storing information. The scheme has a source of information whose output, a message,  is transformed by a coder into a signal acceptable for sending through the channel or storing it in a  memory device. The channel has a noise source that describes errors due to its physical nature. In the output of the channel a decoder transforms the message into the signal suitable for  the receiver. Considering printing one can find the similar organization, shown in Fig. \ref{b1} \hypertarget{b1}. Here the source of information is a digital grayscale image. By halftoning the image is transformed into a binary one, that is input of  the printing binary channel. This channel consists of  "ink"  and "paper".  Output is a  printed image addressed for the receiver. The receiver plays an important role, his features are very complicated and we will ignore him. So the introduced scheme is not perfect but we think this scheme may be sufficient to model printing process.

The binary channel transmits  strings of  0 and 1  generated by the printing device
 whose main elements are "ink" and "paper". The terms "ink" and "paper" are conventional. They can be a usual ink and paper, when the image is presented by an usual way. Also, one can find electronic paper or inks. Anyway the methods of reproducing halftone or colored image have the similar features.

\section{Binary noise}
The binary channel has a binary noise. To model numerous sources of noise in printing process by the binary channel a set of controlled gates can be suitable.

Because of the noise  zeros and ones can be transmitted incorrectly from input to output. It results in information loss and decreasing capacity of the communication channel. For  printing it means a degradation of the image quality. Two main types of noise are known as bit-flip $0\to 1,$ $1\to 0$,  and erasing $\{0;1\}\to 1$. They occur with some probability $p$.  For example, if  the bit-flip error has $p=1/2$ then the channel capacity is zero and information can't be transmitted.
There are numerous noise sources in printing process. For example, consider  a printing head, or more generally  the printing devices. They have to produce an ink dot of a fixed size in a given position. Let  introduce a simple vibration, it affects on the printing  head and  the head makes errors. Then one can find an unnecessary dot, a dot in the unwanted position  also the dot size may be increased and so on.

The observations tell that in printing process the noise can be described mathematically by  conditional operations C-U, known in the computer literature \cite{Deutsch}
. The C-U gate or controlled gate,  has two inputs and two outputs named as control and  target. The gate works as follows. If and only if the control bit is 1, then  operation $\mathbf{U}$ is applied to target.  Let $x\in\{0;1\}$ be a control bit  and $y$ be  a target input, then C-U  maps   $x\otimes y \to  x\otimes \mathbf{U}^{x}y.$  By this way one can describe any noise, particular the vibration of the printing head, assuming $x$ be a random  string of 0 and 1 and $\mathbf{U}$ be an appropriate operation.  In more general case both the control and target signal can be \emph{n-d} arrays. If we wish to describe  processing image then $n=2$ and it needs 2d-binary noise.

\begin{figure}
  \includegraphics[width=18cm]{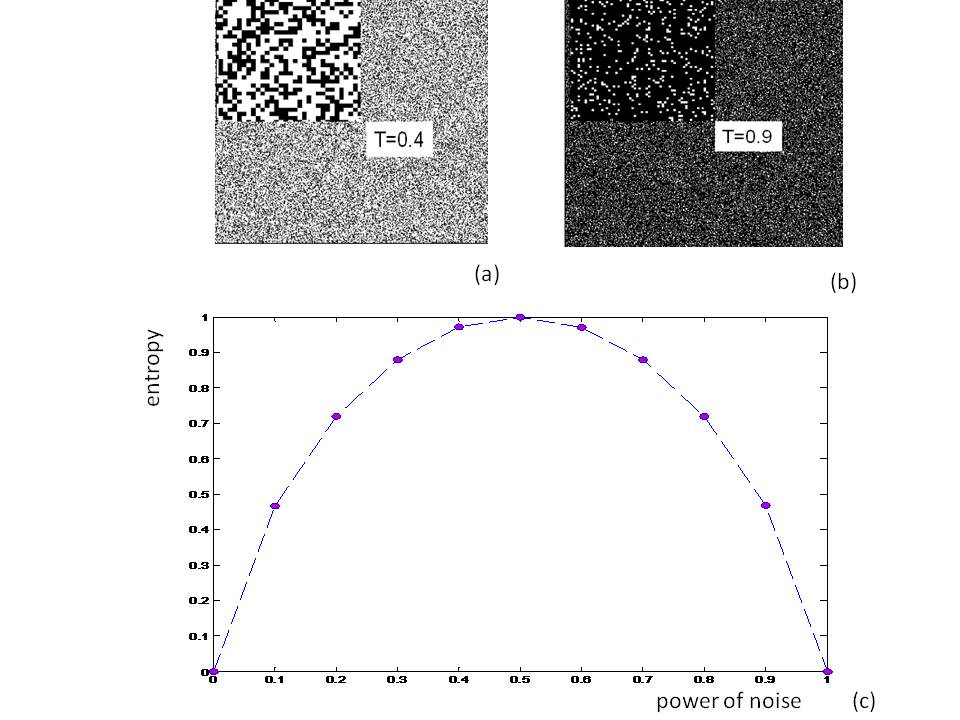}\\
  \caption{Binary noise and its properties.   Random binary pattern obtained by a threshold halftoning vs threshold T, that is power of noise (a) and (b), entropy of binary noise (c).
   }\label{b2}
\hyperlink{b2}{W}
\end{figure}

We generated  binary noise by a simple way with the help of  threshold halftoning.  Let $\mathbf{r}$ be a random matrix, a 8-bit grayscale image. Let the probability distribution function of lightness  $\mathbf{r}[m,n]=0,1,..,255$ be uniform. Then binary matrix  is obtained  like that
	 \begin{eqnarray*}
    \mathbf{v}[m,n]=\begin{cases}
                             1, &\text{if  $\mathbf{r}[m,n]< T$;}\\
                             0, &\text{if  $\mathbf{r}[m,n]\geq  T$,}
                             \end{cases}
  \end{eqnarray*}
where $T$  is a threshold. It follows from this equation  that in matrix $\mathbf{v}$  the number of ones is proportional to threshold $T$ so we will name $T$   power of binary noise.  Fig. \ref{b2} \hypertarget{b2} shows some results on  binary noise and its entropy, presented as functions of $T$.  Perceptually  noise refers to lightness of the image. The image looks more black when power increases. Due to random matrix $ \mathbf{r}$ entropy of binary noise  is a random function of $T$  and it needs averaging.  In Fig. \ref{b2} (c)  entropy has been calculated by averaging over 1024 realizations of $\mathbf{r}$.  Note, entropy achieves its  maximum at $T=0.5$  because this threshold results in  matrix $\mathbf{v}$ that  has zeros and ones  generated uniformly with probability close to 1/2.

\begin{figure}
  \includegraphics[width=18cm]{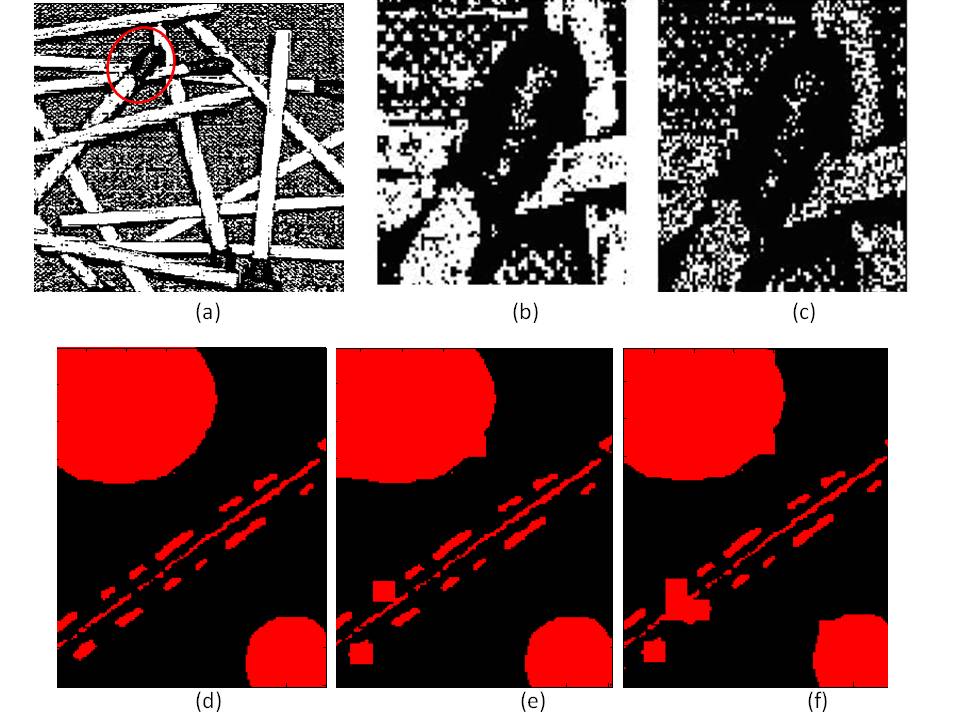}\\
  \caption{Erasing noise.}
\label{b3}
\hyperlink{b3}{W}
\end{figure}

 Erasing noise $\{0,1\}\to 1$ can be modeled by a C-U gate with 2d input. Let the control signal be the binary noise $\mathbf{v}$ and let input target signal be a halftoned image $\mathbf{g}$. Then $x=\mathbf{v}[m,n]$, $y=\mathbf{g}[m,n]$  and the C-U  gate    accomplishes the task, if  $\mathbf{U}^{x}y=x\vee y$, where $\vee$ is disjunction. This noise increases number of the black dots at printed image. An example of erasing is shown in \ref{b3} (a,b,c),  noise results in \hypertarget{b3}distortion of contour.  A more exact erasing noise can process a block $K_{g}$ of binary image. Let  one of the pixels,  $\mathbf{g}[m,n]\in K_{g}$,  be transformed into 1 by erasing noise, if and only if the center pixel of the block  $\mathbf{g}[m_{c},n_{c}]=1$. For this case input of the C-U gate is   $x=\mathbf{v}[m,n]\otimes \mathbf{g}[m_{c},n_{c}] $ and  $y=\mathbf{g}[m,n]$. The gate accomplishes the task, if   $\mathbf{U}^{x}y=\mathbf{v}[m,n]\wedge  \mathbf{g}[m_{c},n_{c}] \vee y$, where $\wedge$ is conjunction.  This noise may lead to blooming and smearing of image as it is shown in Fig. \ref{b3}(d,e,f). 

\section{Robustness to noise }

By considering distortion  of a halftoned  image due to noise one can learn about its robustness to noise. If the image is created by a halftoning algorithm, then one can study robustness of the algorithm.

Let a halftoned image $\mathbf{g}$ be transmitted trough a channel with binary noise $\mathbf{v}$ of power $T$ and $\mathbf{g}'$ be the output noisy image. To describe distinction between $\mathbf{g, v, g'}$ we introduce two measures of distortion, euclidian distance and relative entropy. Both measures maps pairs of images given by matrices $\mathbf{A}$ and $\mathbf{B}$ into $[0,\infty)$.

\begin{figure}
  \includegraphics[width=18cm]{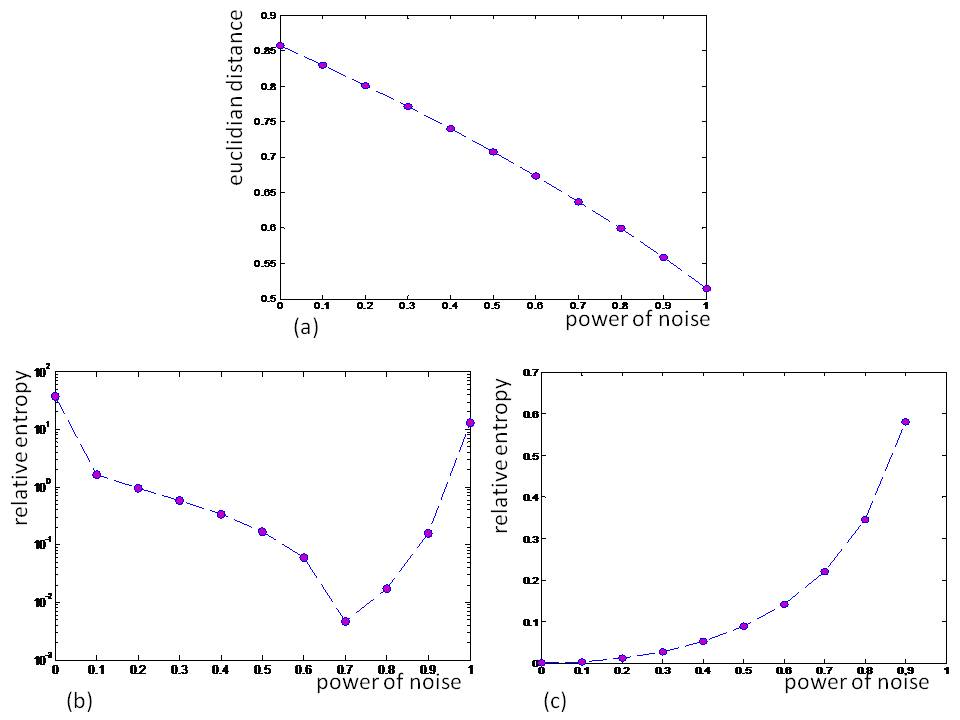}\\
  \caption{Euclidian distance and relative entropy vs power of noise. Euclidian distance between halftone image and binary noise (a), relative entropy between halftone image and binary noise (b), relative entropy between  halftone image and noisy image (c).
   }\label{b4}
\hyperlink{b4}{W}
\end{figure}

Euclidian distance between two matrices is denoted by $e(\mathbf{A,B})=\{(1/n)\sum_{m,n}(\mathbf{A}[m,n]-\mathbf{B}[m,n])^{2}\}^{1/2}$, where $n$ is  the number of elements of matrices. This is an average distance between matrices. It describes how much images $\mathbf{A}$ and $\mathbf{B}$ are close,  the more  euclidian distance the more discrepancy  between images. But this is a mathematical measure and it may be not in accordance with visual perception.
To examine a halftoned image $g$ obtained by D-algorithm introduced in our work \cite{YaD} has been chosen.  We calculated euclidian distance $e(\mathbf{v,g})$  between binary noise and halftone image  and found that  it increases monotonically with  power noise $T$, Fig. \ref{b4} (a). \hypertarget{b4} It means that the binary noise of maximally power is maximally closer to image $\mathbf{g}$.  It  doesn't correspond to visual perception and this is one of the reasons to consider relative entropy.

In statistics the relative entropy known also as Kulback-Leibler entropy  is used to establish the difference between two distributions  of a random variable. For images $\mathbf{A}$ and $\mathbf{B}$  it  is defined as $Q(p_{A}||p_{B})=\sum_{i}p_{A}[i](\log p_{A}[i]-\log p_{B}[i])$, where $p_{A}[i]$, $ p_{B}[i])$ are histograms of lightness $i=0,1,..$. The relative entropy is not negative and   $Q(p_{A}||p_{B})=0$ if and only if $p_{A}=p_{B}$. The relative entropy  is  not a metric because it is in general not symmetric $Q(p_{A}||p_{B})\neq Q(p_{B}||p_{A})$   and does't satisfy the triangle inequality. It is nevertheless  often used in image processing, particular in steganography  and digital watermarking \cite{stg}.

We  calculated the relative entropy $Q(\mathbf{g}||\mathbf{v})$ between the considered above halftone  image $g$ and binary noise. It is shown in Fig. \ref{b4} (b) versus the noise power $T$.  As it is seen  the function $Q(\mathbf{g}||\mathbf{v})$   has its minimum at   $T=0.7$ and increases if  power noise moves off  0.7.
The minimum tells that  histograms of  $\mathbf{g}$ and  $\mathbf{v}$ are very close, if $T=0.7$.  This is a mathematical fact and it is not accordance with the human vision. We think the reason is  that these images have a quite different semantic information that can't properly be described by relative entropy.  Clear, that from the semantic point of view two images $\mathbf{g}$ and $\mathbf{g'}$ are more ''closer'', than $\mathbf{g}$ and noise $\mathbf{v}$. Indeed, we calculated the relative entropy $Q(\mathbf{g}||\mathbf{g}')$ that  increases monotonically with the noise power $T$ (Fig. \ref{b4} (c)).  This is an evident fact because noise introduces a difference in histograms and what is more important this is in accordance with human vision.

Taking into account  observation about semantic information we will use the relative entropy for comparative studies of robustness to noise between halftoning algorithms. This choice is also based on the next statement. \emph{The relative entropy can describe an expert estimation of the quality of the image. }

Let $\mathbf{g}$ and $\mathbf{g}'$ be input and output image of the binary noisy channel and $\mathbf{g}$ be obtained by a halftoning algorithm $\mathbf{a}$. We will name the algorithm $\mathbf{a}$ $\epsilon$-robust to noise, if relative entropy $Q(\mathbf{g}||\mathbf{g'})\leq \epsilon$. The algorithm is perfect, if  $Q(\mathbf{g}||\mathbf{g'})=0$.
The introduced definition tells that the smaller $\epsilon$ is, or the closer the histograms of $\mathbf{g}$ and $\mathbf{g'}$ are, the greater robustness of algorithm is. To better understand what is meant by $\epsilon$-robustness consider that an expert examines the  quality by viewing images $\mathbf{g}$ and $\mathbf{g}'$. Let the expert have to answer the question, is the given image  $\mathbf{g}$ or $\mathbf{g}'$. When $\epsilon$ is small he can make two types of errors.  So the expert can decide that the given image is $\mathbf{g}$, when it fact it is $\mathbf{g'}$ and vice versa. This detection scheme is well known in statistical signal detection and estimation \cite{Detec}. Let probability of the expert errors be $\alpha$ and $\beta$.  Then, if only $\mathbf{g}$ and, if only $\mathbf{g'}$ is shown to expert, the relative entropy  between his decisions is $Q(\alpha||\beta)$.  Because the expert with his decisions is a type of processing and processing can't increase relative entropy for $\epsilon$-robust algorithm one finds
\begin{eqnarray*}
Q(\alpha||\beta)\leq Q(\mathbf{g}||\mathbf{g'})\leq \epsilon.
\end{eqnarray*}
As it follows from this equation $\epsilon$-robustness of algorithm refers to errors of the expert, that tests the image quality.

The relative entropy can be used to compare robustness to noise between halftoning algorithms. Let
$\mathbf{g}_{k}$  be an image obtained by a halftoning algorithm $\mathbf{a}_{k}, k=1,2,\dots$   and  let $\mathbf{g'}_{k}$ be its noisy version.  One can calculate the relative entropies
$q_{k}=Q(\mathbf{g_{k}}||\mathbf{g'_{k}})$.  Algorithm $\mathbf{a}_{k}$ is more robust than $\mathbf{a}_{t}$, if $q_{k}\leq q_{t}$.  In practice this inequality may hold in some region of parameters and for certain types of images because there are not ideal halftoning algorithms.

\begin{figure}
  \includegraphics[width=18cm]{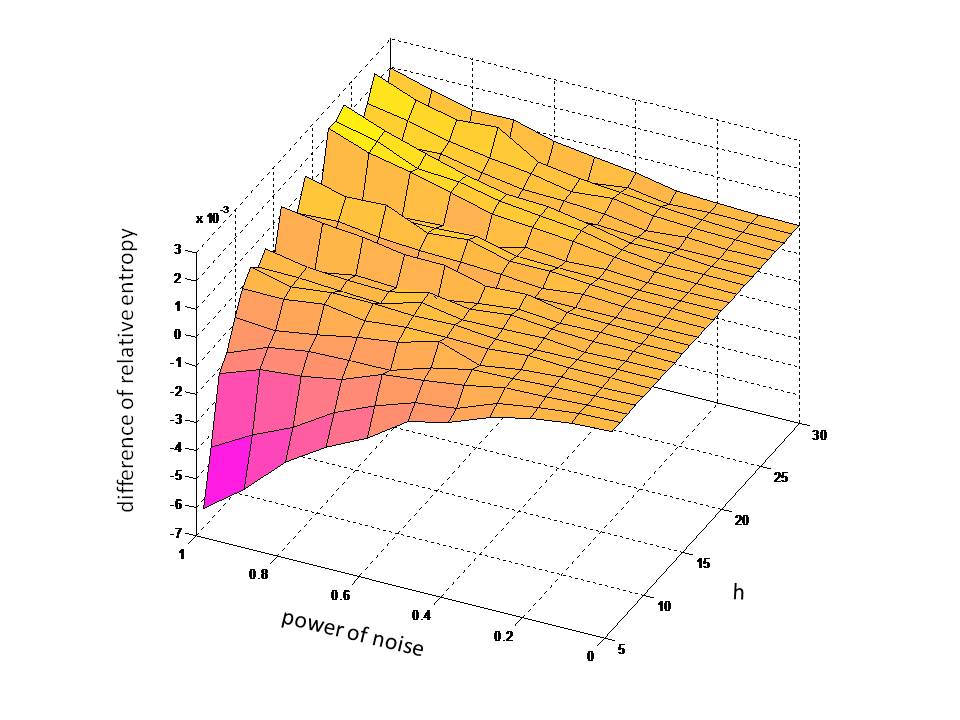}\\
  \caption{Difference between relative entropy of DITHER and D-algorithm.  If the difference is positive  immunity of D-algorithm  to noise is  greater and vice versa. }\label{b5}
\hyperlink{b5}{W}
\end{figure}

\begin{figure}
  \includegraphics[width=18cm]{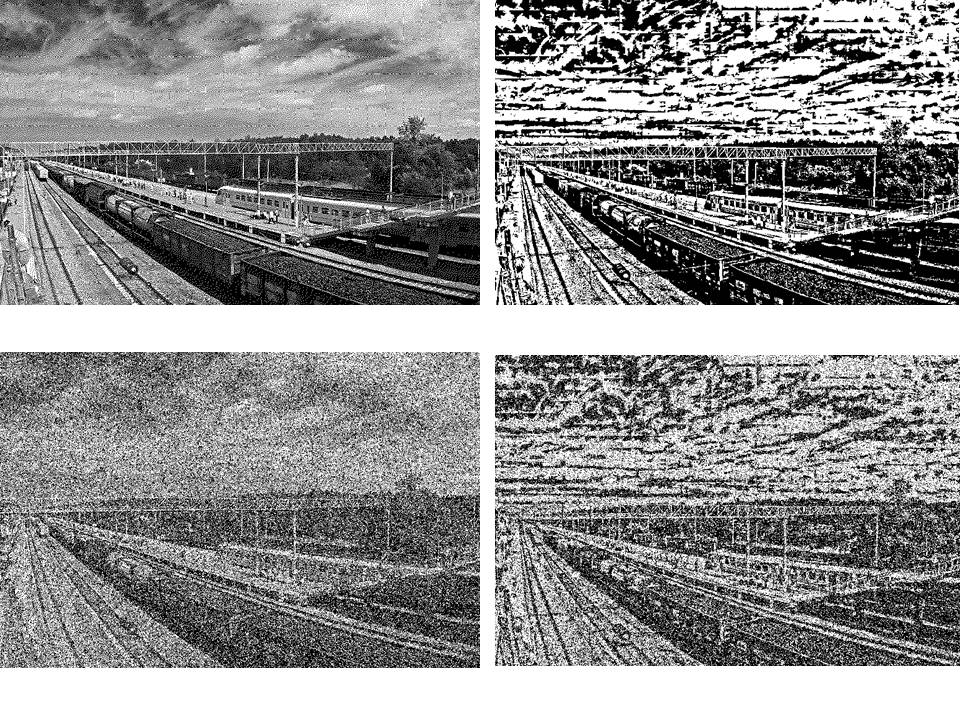}\\
  \caption{Distortion introduced by bit-flip noise with power  $T=0.2$. Halftone images created by DITHER (left) and D-algorithm (right), $h=19$, their noisy versions at the bottom.
     }\label{b6}
\hyperlink{b6}{W}
\end{figure}

\begin{figure}
  \includegraphics[width=18cm]{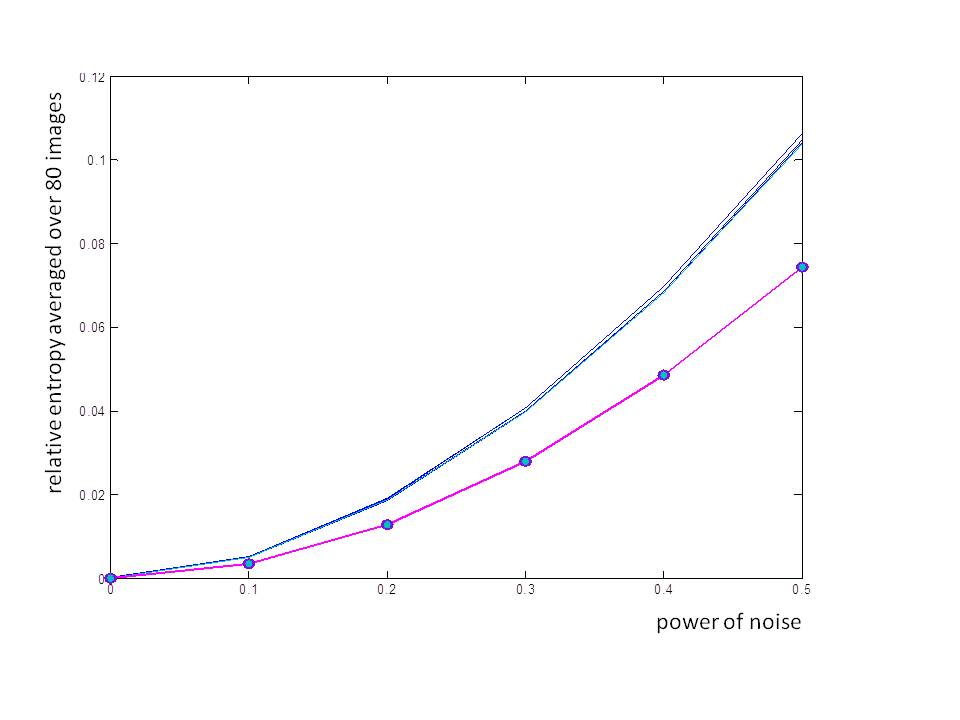}\\
  \caption{Relative entropy overaged over 80 images. Seven halftoning algorithms DOTDIF, ERRDIF, LAU, ULICHNEY, CDOT1, CDOT2, BAYER show very close results, blue curves. The eighth algorithm D is distinguished, red curve, its relative entropy is  least.
     }\label{b7}
\hyperlink{b7}{W}
\end{figure}

As example we studied some algorithms. First is DITHER, it
 is based on the the Floyd-Steinberg error diffusion method and can be evaluated as \texttt{dither(I)} in MATLAB. Second is  D-algorithm, it has  an input  parameter $h$ that denotes  a  grayscale  $h\times h$ block processed \cite{YaD}.  We calculated the relative entropy between  DITHER and D-algorithm,  $q_{1}$ and $q_{2}$ for bit-flip noise.   Fig. \ref{b5}  \hypertarget{b5}  shows difference $q_{1}-q_{2}$ as  function of the power noise $T$ and size block $h$. It follows that there is a range, $h\geq 25$, where $q_{2}< q_{1}$. It this range D-algorithm is more robust against noise.  Fig. \ref{b6}  \hypertarget{b6}  illustrates distortions introduced by noise in halftone images obtained by D and DITHER techniques for  $T=0.2$ and $h=19, q_{2}< q_{1} $. One can see that  distortions are more noticeable   in the DITHER techniques, that is in agreement with  mathematical results. The results presented at Fig. \ref{b5} and Fig. \ref{b6} are valid for an image only and it needs a large set of images for investigation. Clear, that a given halftoning algorithm can't processes properly all possible images. Then there is a problem to find a suitable class of images. One of the solutions is presented at Fig. \ref{b7}. It shows relative entropy of eight halftining algorithms versus power of noise. We used seven original halhtoning algorithms
 \texttt{DOTDIF, ERRDIF, LAU, ULICHNEY, CDOT1, CDOT2, BAYER}  from \cite{1} and D algorithm.
 Each point at Fig. \ref{b7}  \hypertarget{b7}  is averaged over 80 images from the famous collection Caprichos by Goya.
 Indeed, there is a monotonic behavior that tells that relative entropy of D algorithm is the least. It means, for example, that to print Caprichos D algorithm may by more preferable.



\newpage
\newpage


\end{document}